\documentstyle [11pt,epsf]{article}

\begin{document}
%
%
\newcommand{\Abs}[1]{|#1|}
\newcommand{\EqRef}[1]{(\ref{eqn:#1})}
\newcommand{\FigRef}[1]{fig.~\ref{fig:#1}}
\newcommand{\Abstract}[1]{\small
   \begin{quote}
      \noindent
      {\bf Abstract - }{#1}
   \end{quote}
    }
\newcommand{\FigCap}[2]{
\ \\
   \noindent
   Figure~#1:~#2
\\
   }
%
%
%
%
\title{Approximate zeta functions for the Sinai
billiard and related systems}
\author{Per Dahlqvist \\
Mechanics Department \\
Royal Institue of Technology, S-100 44 Stockholm, Sweden\\[0.5cm]
}
\date{}
\maketitle

\ \\ \ \\

%
\Abstract{We discuss zeta functions, and traces of the associated weighted
evolution operators for intermittent Hamiltonian systems in general and for
the Sinai billiard in particular. The intermittency of this billiard is
utilized so that the zeta functions may be approximately expressed in terms
of the probability distribution of laminar lenghts.
In particular we study a one-parameter family of weights. Depending on the
parameter the trace can be dominated by branch cuts in the zeta function
or by isolated zeros. In the former case the time dependence of the trace
is dominated by a  powerlaw and in the latter case by an exponential.
A phase transition occurs
when the leading zero collides with a branch cut.
The family considered is relevant for the calculation of resonance spectra,
semiclassical spectra and topological entropy.}

\ \\
\section{Introduction}

The space of Hamiltonian systems is framed by two extreme cases -
integrable systems and strongly chaotic systems, the latter in the sense of
Axiom-A. Both cases may be successfully
{\em solved}, but the methods of solution look entirely different.
For integrable systems {\em solving} means {\em finding} the
constants of motion. In the Axiom-A case one rather aims at a  statistic
description of the motion - one calculates different kinds of entropies,
fractal dimensions, resonance spectra etc.
To this end {\em cycle expansions} of zeta
functions have shown very successful \cite{AAC,CE,Rosen}.
So far, the success of this approach
has been limited to the Axiom-A case. The reason is that Axiom-A ensures
nice analytic features of the zeta functions involved
\cite{Polli,Rue1,Rue2,Rugh}, whereas different
kinds of singularities emerges when departing from Axiom-A \cite{PDreson}.
If the system is not
Axiom-A, but still ergodic, it is
generally intermittent (there are cycles
with arbitrary small Lyapunov exponents) and lacks a simple symbolic
dynamics. This is e.g. the case for most bound ergodic systems \cite{PD1}.
It is of great interest to extend the study of zeta
functions to this case.
The resulting theory would be an analogue of perturbation
theory for almost integrable systems.

Such a theory would be interesting  not only for ergodic systems.
The motion in a chaotic sea of a generic mixed system
is in general intermittent. This is because the trajectory
may be trapped inside {\em cantori} surrounding the stable islands, where
it exhibits quasi-integrable motion.

The cycle expansion for a system with a finite subshift symbolic dynamics
consists of two parts \cite{AAC}, a {\em fundamental part}, giving the gross
structure of the spectrum, and {\em curvature corrections}.

When the symbolic dynamics is an infinite subshift, there is
no similar division.
One goal of this paper is to establish a fundamental part for this case.
One can think of two ways out of the dilemma. One can
define succesive finite subshifts, approaching the infinite subshift,
by utilizing
a {\em pruning front} \cite{skel,Kai}.
This mean that one persists on a periodic orbit
description.
We will attack the problem from the opposite end
and completely abandon the periodic orbits.
We will thus make extensive use of our systems being
ergodic and Hamiltonian, so that
the invariant density is known a priori. One of the morals to be taught
is that there is no reason to find a lot of periodic orbits in order
to compute something you already know in advance.
The idea to use probabilities rather than periodic orbits has been used in a
similar way in ref. \cite{Gregor}.

In section 2 we develope most of the theoretical apparatus needed. The
exposition will be rather brief as most of the material can be found
elsewhere.
In section 3, we apply these ideas to the Sinai billiard
and formulate an approximate zeta function.
In section 4 we study this zeta function numerically and
compare with
periodic orbit theory.

\section{Theory}

Much of the theoretical work on these matters are centred around the
evolution operator. It describes the evolution of a phase space density
$\Phi(x)$
\begin{equation}
{\cal L}_w^t  \Phi(x)=\int w(x,t)\delta (x-f^t(y))\Phi(y)dy   \ \ .
\end{equation}

The phase space point $x$ is taken by the flow to $f^t(x)$ during time
$t$. $w(x,t)$ is a weight associated with a trajectory starting at $x$
and evolved during time $t$. It is multiplicative along the flow, that
is $w(x,t_1+t_2)=w(x,t_1)w(f^{t_1}(x),t_2)$.

We emphasize that $t$ is a continous variable as we are studying flows
and not maps.

\subsection{Trace formulas and zeta functions}

Our main concern in this paper is to compute the trace of the evolution
operator, that is, the sum of its eigenvalues.
The material in this subsection may be found e.g. in refs. \cite{AAC,flows}

The trace may be written as a sum over the
periodic orbits in the system
\begin{eqnarray}
tr {\cal L}_w^t  =\int w(x,t) \delta (x-f^t(x))dx= \nonumber \\
\sum_p T_p \sum_{n=1}^{\infty} w_p^n \frac{\delta(t-nT_p)}
{\Abs{det(1-M_p^n)}}  \ \ ,
\label{eqn:tracedef}
\end{eqnarray}
where $n$ is the number of repetitions of primitive orbit $p$, having period
$T_{p}$, and  $M_{p}$ is the Jacobian of the Poincar\'{e} map. $w_p$ is the
weight associated with cycle $p$.

The trace is written as the
Fourier transform of the logarithmic derivative
of a {\em zeta function}
\begin{equation}
tr {\cal L}_w^t = \frac{1}{2\pi i}
\int_{-\infty}^{\infty} e^{ikt}\frac{Z_w'(k)}{Z_w(k)}dk   \ \ .
\label{eqn:tracefour}
\end{equation}
We restrict ourselves to systems with two degree's of freedom
for which all periodic orbits are isolated and unstable.
The zeta function then
reads
\begin{equation}
     Z_w(k)=\prod_{p}\prod_{m=0}^{\infty}
          \left(1-w_p \frac{e^{-ikT_{p}}}
    {\Abs{\Lambda_{p}} \Lambda_{p}^{m}}\right)^{m+1}
          \ \ ,                       \label{eqn:Zw}
\end{equation}
where $\Lambda_{p}$ is the expanding
eigenvalue of $M_p$.

Putting $w=1$ we obtain a zeta function
whose zeros yields the so called {\em resonance spectrum} or
spectrum of correlation exponents,
\begin{equation}
     Z(k)=\prod_{p}\prod_{m=0}^{\infty}
          \left(1- \frac{e^{-ikT_{p}}}
    {\Abs{\Lambda_{p}} \Lambda_{p}^{m}}\right)^{m+1}
          \ \ .                       \label{eqn:Zcl}
\end{equation}
The leading zero, $k_0$, is the escape rate, which is zero for a bound
system.

By using different weights $w$ one can probe different properties
of the flow.

The topological entropy is obtained by considering the weight
$\omega = \Abs{\Lambda(x,t)}$.
$\Lambda(x,t)$ is the expanding eigenvalue
of the Jacobian transverse to the flow.  It is only
approximately
multiplicative along the flow but it is possible to modify it slightly so as
to become exactly multiplicative \cite{Gabor}.
However,
this subtle difference is of no
importance for our purposes.
The trace formula then becomes
\begin{equation}
tr {\cal L}_{top}^t  \approx
\sum_p T_p \sum_{n=1}^{\infty} \delta(t-nT_p)
\ \ .
\end{equation}
The leading zero  lies on the negative imaginary axis.
The topological entropy is then $h=ik_0$, the asymptotic behaviour of the trace
is
\begin{equation}
\sum_p T_p \sum_{n=1}^{\infty}  \delta(t-nT_p)
\rightarrow e^{ht}
\end{equation}
so that the number of cycles with periods less than $t$ is
$\sim e^{ht}/ht$, which is the familiar result.

The semiclassical case is obtained by considering the weight
$w=\sqrt{\Abs{\Lambda(x,t)}}$.  We now get the following zeta function,
\begin{equation}
     Z_{sc}(k)=\prod_{p}\prod_{m=0}^{\infty}
          \left(1- \frac{e^{-ikT_{p}}}
    {\sqrt{\Abs{\Lambda_{p}}} \Lambda_{p}^{m}}\right)^{m+1}
          \ \ .                       \label{eqn:Zsc}
\end{equation}
This zeta function is called {\em the quantum Fredholm determinant}
\cite{Rosen} and is equivalent to the
Gutzwiller-Voros \cite{Gut0,Vor} zeta function in the semiclassical limit, and
have indeed
nicer analytical features.
We have left out the Maslov indices
but it is possible to account for them in the weight as well.


\subsection{Symmetry decomposition of $Z(k)$}

Many systems possess finite symmetries. This enables
a factorization of the zeta function. This was first shown for the
semiclassical case in refs. \cite{Rob,L} and  for a more general setting
in ref. \cite{CEsym}.
The zeta function $Z_w(k)$ in \EqRef{Zw} is then written as a
product $Z_w(k)=\prod_{r} Z_{w,r}(k)$
over $r$, the irreducible representations of the
symmetry group $G$, and
each $Z_{w,r}$ is given by
\begin{equation}
     Z_{w,r}(k)=\prod_{p}\prod_{m=0}^{\infty}
          \left(1-\chi_{r}(g_{p})w_p
      \frac{e^{-iT_pk}}
            {\Abs{\Lambda_{p}}\Lambda_{p}^{m}}\right)^{m+1}
          \ \ , \label{eqn:Zr}
\end{equation}
where $p$ runs over all prime cycles in the fundamental domain.
$\chi_{r}(g_{p})$ is the group character for the irreducible
representation $r$, and $g_{p}$ is the group element obtained as the
product of group factors associated with the reflections of the domain walls.

Throughout the paper we will restrict ourselves to one-dimensional irreducible
representations of the group.
Periodic orbits running along symmetry lines must be given a special
treatment. We will discuss this when the result is needed.

In the quantum case the different irreducible representations corresponds
to the different symmetry classes of the wave functions. But in the general
classical case it is harder to give a comprehensible interpretation. The
important thing to keep in mind in the following is that the leading zero is
always in the symmetric representation.

\subsection{Zeta functions in the BER approximation}

In ref. \cite{PDreson} an approximate expression for the zeta function
is given for intermittent, ergodic Hamiltonian systems. The idea
is based on a paper
by Baladi, Eckmann and Ruelle \cite{BER} and it is therefore referred to
as the BER approximation.

In an intermittent system there are two, more or less, distinct
phases; one regular and one irregular
(chaotic).
Call the consecutive instants when the system enters the regular phase
$\{ t_i \}$ and consider the intervals  $I_i=[t_{i-1},t_i]$.
Provided the chaotic phase is {\em chaotic enough},
the motions in different intervals are nearly mutually independent.
This is called {\em assumption A}.
In particular, the lengths of these intervals
$\Delta_i=t_i - t_{i-1}$ are mutually independent
and $\Delta$ may be considered as a stochastic variable
with probability distribution $p(\Delta)$.
Under this assumtion, the zeta functions may be expressed in terms
of the Fourier transform of $p(\Delta)$
\begin{equation}
Z(k)\approx \hat{Z}(k)
\equiv 1-\int_{0}^{\infty}e^{-ik\Delta}p(\Delta)d\Delta
\end{equation}
We use the $\hat{}$ - symbol to denote all quantities
in the BER approximation.

This approximation is not restricted to unit weight.
If the weight associated with one interval
is a function of the length of the interval $\Delta$ only, then
\begin{equation}
\hat{Z}_w(k)=1-\int_0^{\infty} e^{-ik\Delta } w(\Delta)p(\Delta)d\Delta
\ \ .
\label{eqn:hatZw}
\end{equation}
This is one of the central ideas of this paper.
To show this, one just repeats the calculation of ref. \cite{PDreson},
squeezing
in the weight at the appropriate places. As it is straightforward to do
we will not perform it here.
The assumption above, that the weight is a function of $\Delta$ only,
is quite reasonable for many systems. For the Sinai billiard, which
we will study soon, the expression has to be slightly generalized.

\begin{figure}[p]
\vspace{15cm}
\caption{Three equivalent representations of the Sinai billiard.
a) The fundamental domain. b) The original Sinai billiard with definitions
of the variables $\phi$ and $\alpha$. c) The unfolded system with a sample
of orbits which are periodic of period one in the fundamental domain.}
\end{figure}

\section{The Sinai billiard}

The  purpose of this section is to
apply the BER approximation, that is formulate $\hat{Z}_{w,r}(k)$, for the
Sinai billiard \cite{Sin} and compare with cycle expansions.
The billiard consist of a unit square with a
scattering disk, having radius $R$;
$0<2R \leq 1$, centered on its midpoint, cf. fig. 1b.
The Sinai billiard is fairly simple but have all the
typical features for bound
chaotic system; it is intermittent and lacks a simple
symbolic dynamics. Moreover, it has marginally stable orbits.
However, a trajectory in the hyperbolic phase (hyperbolic in the sense that
all cycles there are unstable and isolated)
will never end up on a
marginally
stable orbit, it can only come
arbitrary close to.
This means that we can
separate out
this marginal phase. That is, the integral \EqRef{tracedef}
is only performed
over the hyperbolic part of phase space, thus excluding a region
of measure zero, so that the
sum in eq. \EqRef{tracedef} runs only over isolated
periodic orbit.

This separation cannot be performed in the quantum case due to an extra
contribution from the marginal orbits only disappearing in the limit
$E \rightarrow \infty$. The study of the quantum case will be postponed
to a forthcoming paper.

The trajectory of the Sinai billiard consists of laminar intervals,
when bouncing between the straight sections, interrupted by scatterings
on the central disk. The Sinai billiard seems therefore ideally suited
for the BER approximation. When the disk radius is small,
the memory of the previous laminar interval should be almost
completely lost. We will not discuss corrections to the BER approximation
arising from the correlations between laminar intervals. We simply assume
that they are small when the disk radius is small.
The length of
the chaotic interval is infinitely short so that $\Delta$ is simply the length
of the trajectory between two
disk scatterings. For big radii, there is another source of intermittency
\cite{Macht,Bouch}, the trajectory may be
trapped between the disk and the straight
section. For these reasons we will focuse on the limit of small disk radii.

\subsection{Symbolic dynamics}

First we will define a symbolic dynamics for the Sinai billiard.
The reason is twofold.
First we need it for the application of the BER approximation. Secondly
we will use it for finding the periodic orbits of the system,
which we will need to test the results obtained from
the BER approximation.

We let the
disk be our Poincar\'{e} surface of section. The canonical variables
are
$(\xi,p_{\xi})=(R\phi,\sqrt{2E}sin\alpha )$ where the angles
$\phi$: $0< \phi< 2\pi $ and $\alpha$: $-\frac{\pi}{2}<\alpha<\frac{\pi}{2}$
are defined in fig. 1b.
The area preserving map
$(\xi,p_{\xi}) \mapsto (\xi,p_{\xi})$ has uniform invariant density.

We now want to define a symbolic dynamics,
such that each iterate of the map corresponds to one symbol, that is,
each symbol corresponds
to one laminar interval.

To do this we define two equivalent
billiard systems, derived from the original system.
The original billiard has the
symmetry group $G=C_{4v}$, \cite{Hamm}.
This group has eight elements in five conjugacy classes, see table
1.

The first derived system is the fundamental
domain, or the desymmetrized system, see fig 1a.
This system has, by definition, no
symmetry, except for time reversal.

To obtain the second derived system we go in the opposite direction.
We unfold the billiard, see fig 1c, into a regular lattice of disks. Apart
from the group $C_{4v}$  this system
has also a discrete translational symmetry.

We note that any trajectory
in the unfolded billiard may be encoded by a sequence of column
vectors $\ldots Q_{i-1} Q_i Q_{i+1}\ldots$
\begin{eqnarray}
Q_i \in \Omega \nonumber \\
\Omega : \{  \left( \begin{array}{c} n_x \\ n_y \end{array} \right)
; n_x,n_y \in Z, gcd(n_x,n_y)=1 \}
\end{eqnarray}
where $gcd(n_x,n_y)$ stands for the greatest common divisor of
$n_x$ and $n_y$.
The number $n_x$ ($n_y$) simply tells the number of disk sites
the segment of
the trajectory has travelled in the horisontal (vertical) direction.
It is easy to see that non comprime $Q$'s cannot
be realised.

A general trajectory corresponds to a semi infinite string of $Q$'s.
We will focuse on
periodic strings
$\overline{Q_{1} Q_2 \ldots Q_{n}}$.
Such a string uniquely determines
the trajectory in the unfolded billiard
up to a translation which means that
the corresponding trajectory is completely
determined in the original billiard, and is indeed periodic there.
Of course, some trajectories defined in this way
would need to go through disks. The corresponding
symbol sequences are then said to be {\em pruned}.
But when the disk radius is small we expect this pruning to apply
only to those orbits with some {\em long} segment(s). More exactly
if all the $Q$'s have $\sqrt{n_x^2+n_y^2}  \ll 1/R$ we expect the
pruning to be neglible. This is very convienient for computing puposes
since a large fraction
of the generated symbol sequences will be realised as true orbits.

We observe that a symmetry transformation of the orbit corresponds to
a transformation $Q_i \mapsto g Q_i$ where g is an element of $C_{4,v}$
represented by a $2 \times 2$ matrix, see table 1.

Our goal is to encode cycles in the fundamental domain.
Due to symmetry, several
cycles of $Q$'s correspond to one and the same cycle
in the fundamental domain. We therefore want to translate the cyclic symbol
string $\overline{Q_{1} Q_2 \ldots Q_{n}}$  to another string
$\overline{s_{1} s_2 \ldots s_{n}}$ where the $s$'s are taken from another
alphabet. The translation must
fulfill the following conditions:

\begin{enumerate}
\item The translation is unique.
\item The sequence $\overline{s_{1} s_2 \ldots s_{n}}$ is invariant
under a symmetry operation on $\overline{Q_{1} Q_2 \ldots Q_{n}}$ .
\item A cyclic shift on the $Q$'s corresponds to a cyclic shift on the
$s$'s.
\end{enumerate}
This will be (almost) achieved in the following way.
Let each $s_i$ be an ordered pair $s_i=(q_i,g_i)$
\begin{eqnarray}
g_i \in C_{4,v} \nonumber \\
q_i \in \omega  \\
\omega : \{  \left( \begin{array}{c} n_x \\ n_y \end{array} \right)
; n_x,n_y \in N ,n_x \geq n_y, gcd(n_x,n_y)=1 \}  \nonumber
\end{eqnarray}
The translation is defined by:
\begin{eqnarray}
Q_1 =  g_0 g_1 \; q_1 \nonumber \\
Q_2 = g_0 g_1 g_2 \; q_2 \\
\vdots \nonumber \\
Q_{n} = g_0 g_1 \ldots g_n \; q_n \nonumber \\
\end{eqnarray}
The translation is unique as long as no vector $q_i$ is equal to
\( ( {\tiny \begin{array}{c} 1 \\ 0 \end{array} } )  \)  or
\( ( {\tiny\begin{array}{c} 1 \\ 1 \end{array}} )  \)
(we refer to these as boundary
symbols). We will deal with this problem in a while.
Condition 2 is also fulfilled - a symmetry operation affects only
$g_0$. It is easy to check that also condition 3 is fulfilled (a cyclic
shift also affects $g_0$).

Suppose now that $q_j$ is a boundary vector. If we infer the restriction
that $g_j$ must not be a reflection ($\sigma_{\mid}$, $\sigma_{-}$,
$\sigma_{/}$ or $\sigma_{\setminus}$)
the translation is again unique provided that not {\em all} the
$q$'s are boundary symbols.

Suppose the latter is the case and that we have found {\em one} translation
of the string $\overline{Q_{1} Q_2 \ldots Q_{n}}$, namely:
$\overline{s_1 s_2 \ldots s_n}$.
It is now possible to find another $s$-string corresponding to the same
$Q$-string. This is done in the following way.
Suppose that $q_i$ has the symmetry $\phi_i$, that is,
if \( q_i=( {\tiny\begin{array}{c} 1 \\ 0 \end{array}} )  \)
then $\phi_i=\sigma_{-}$ and if
\( q_i=( {\tiny\begin{array}{c} 1 \\ 1 \end{array}} )  \)
then $\phi_i=\sigma_{/}$.
Then the transformed sequence
\begin{eqnarray}
q_i \mapsto q_i  \nonumber \\
g_i \mapsto \phi_{i-1} g_i \phi_i \; \; \; i\neq 0 \\
g_0 \mapsto g_0 \phi_n   \nonumber \\  \label{eqn:trans} \nonumber
\end{eqnarray}
corresponds to the same $Q$ string.
Remember that the string is periodic  so that e.g. $\phi_0=\phi_n$.
Performing this transformation once
more gives back the original $s$ string. So there remains a problem that an
orbit in the fundamental domain might correspond to two symbol strings and
we have to account for this problem when generating the periodic orbits.
The symbolic dynamics defined is still a good one and the problem above should
be intepreted as a boundary effect.

As we said, the problem arises only if all vectors are on the boundary.
If one is not, and we try to perform the transformation \EqRef{trans}, we see
that at least one $g_i$ corresponding to a boundary $q_i$ would transform to
a reflection and such strings have already been forbidden.

\begin{table}

\[
\begin{array}{|c|c|c|}
\hline
E  &  \left( \begin{array}{cc} 1 & 0 \\ 0 & 1 \end{array} \right) &
\mbox{Identity} \\
\hline
I & \left( \begin{array}{cc} -1 & 0 \\ 0 & -1 \end{array} \right) &
\mbox{Inversion} \\
\hline
R^{+},R^{-} &
\left( \begin{array}{cc} 0 & -1 \\ 1 & 0 \end{array} \right),
\left( \begin{array}{cc} 0 & 1 \\ -1 & 0 \end{array} \right) &
\mbox{Rotation in the positive and negative directions} \\
\hline
\sigma_{|},\sigma_{-} &
\left( \begin{array}{cc} -1 & 0 \\ 0 & 1 \end{array} \right),
\left( \begin{array}{cc} 1 & 0 \\ 0 & -1 \end{array} \right) &
\mbox{Reflection in the vertical and horisontal axis} \\
\hline
\sigma_{/},\sigma_{\setminus} &
\left( \begin{array}{cc} -1 & 0 \\ 0 & 1 \end{array} \right),
\left( \begin{array}{cc} 1 & 0 \\ 0 & -1 \end{array} \right) &
\mbox{Reflections in the diagonals} \\
\hline
\end{array}
\]

\caption{The elements of the group $C_{4,v}$. They are represented by
$2 \times 2$ matrices operating on the column vectors $q_i$ and
$Q_i$.}
\end{table}

The group characters
in eq. \EqRef{Zr} are determined both from the disk bounces and the bounces on
the square walls. They
are easily expressed in therms of the symbol code
\begin{equation}
\chi_r(g_p)=\prod_{i=1}^{n_p} \chi_r(g_i)
\chi_r(\sigma_{-})^{n_{x,i}+n_{y,i}}
 \ \ ,   \label{eqn:symmfac}
\end{equation}
To obtain this we have used the familar property of group characters:
$\chi_r(g_1g_2)=\chi_r(g_1)\chi_r(g_2)$.
where $g_i$, $n_{x,i}$ and $n_{x,i}$ are taken from the symbols string
corresponding to prime orbit $p$.

One may also consider the Sinai billiard with periodic boundary conditions.
One then simply exclude the reflection factor
$\chi_r(\sigma_{-})$ in eq. \EqRef{symmfac}.

\subsection{Free Directions}

For any disk radius $0<2R<1$ there are a finite number of directions
through which a trajectory
may go without ever bouncing on any disk. We call them {\em free directions}.
Consider the direction vector $(x,y)$. A direction
cannot be free if $x/y$ is irrational. So we may take $(x,y)$ as
coprime integers such that $x/y \leq 1$.
The direction $(x,y)$ is found to be free
if
\begin{equation}
2R<\frac{1}{\sqrt{x^2+y^2}}  \ \ .
\end{equation}
To show this, consider an integer lattice of points (or consult
ref
\cite{BerSin}).  We are seeking the lattice point
$(\xi,\eta)$ closest to the line from $(0,0)$ to $(x,y)$ and lying below.
The closest point above this line is then, due to symmetry,
$(x-\xi,y-\eta)$. By the definition of Farey sequences, $\frac{\eta}{\xi}$ is
the preceeder of $\frac{y}{x}$ in the Farey sequence of order $x$. The Farey
theorem then says that $\xi y-\eta x =1$ \cite{numb}. This is nothing but the
area of the parallellogram spanned by the points $(0,0)$, $(\xi,\eta)$, $(x,y)$
and $(x-\xi,y-\eta)$ so that the distance between the point $(\xi,\eta)$ to the
diagonal is $1/\sqrt{x^2+y^2}$ and the result above follows.

The direction $(x,y)$ being free is equivalent with the existence of the
periodic orbit $\overline{s}$ where
\( s=( ( {\tiny\begin{array}{c} x \\ y \end{array}} ),E)  \).

\subsection{The zeta function in the BER approximation}

We begin by considering
the symmetric representation $r=A_1$ (then
$\chi_{r}(g_{p})=1$) and unit weight $w=1$ in the zeta function.
The corresponding indices
will be omitted in the zeta function.

Utilizing the uniformity of the invariant measure \cite{Sin} we can write down
the following expression for the probability distribution $p(\Delta)$
\begin{equation}
p(\Delta) \approx \sum_{q \in \omega} a_q(0) \delta (\Delta-T_q) \ \ ,
\end{equation}
where $a_q(0)$ as given by
\begin{equation}
a_q(0)=\frac{2}{\pi} \int_{\Omega_q}d\phi \; d(sin\;\alpha )
\ \ .
\label{eqn:aqdef}
\end{equation}
$\Omega_q$ is the part of phase space (i.e. the
$\alpha,\phi$-plane) for which the trajectory
hit disk $q$ (in the unfolded system).
We have used the symmetry by letting $q \in \omega$, meaning that we take
only orbits
going out into the first octant into account.
Only the fractions of the disks
\( ( {\tiny \begin{array}{c} 1 \\ 0 \end{array}} )  \)   and
\( ( {\tiny \begin{array}{c} 1 \\ 1 \end{array}})  \)
lying in this octant are included.
The argument (now put to zero) in $a_q(0)$ is introduced for later purposes.

We have made the approximation that
all trajectories to disk $q$ have the same length
$l_q$ corresponding to the time of flight $T_q$ (this will be refined
a little in section 3.5).
This is reasonable when the disk radius
is small. Taking the Fourier transform we get:
\begin{equation}
\hat{Z}(k)= 1-\sum_{q} a_q(0) e^{-ikT_q} \ \ .
 \label{eqn:Zhat}
\end{equation}
Although we use equality sign above we must keep in mind that this result
involves an approximation in addition to
the BER approximation.
\begin{figure}
\epsffile{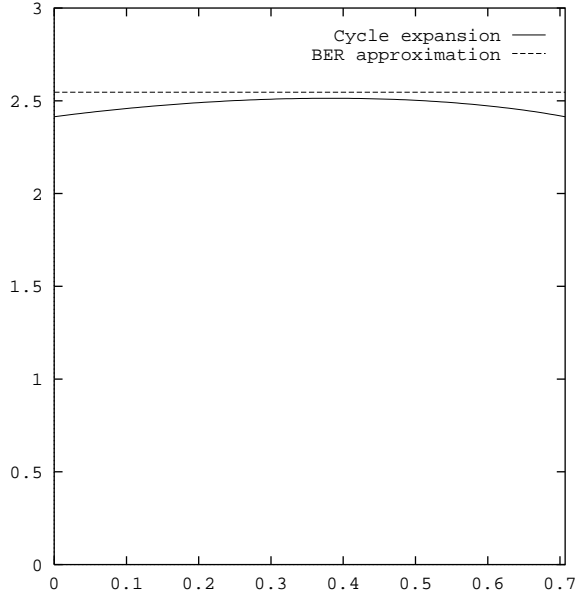}
\caption{Comparison of the prefactors of eqs (22) and (25) plotted
versus $sin\phi$ }
\end{figure}

\subsection{Relation to cycle expansion of $Z(k)$.}

We are now interested in the relations between the
approximate zeta function $\hat{Z}(k)$
and the cycle expansion of the exact zeta function $Z(k)$. First we elaborate
a little more on $\hat{Z}(k)$.

It is convenient to write $a_q(0)=\hat{\eta}_q \frac{8R}{\pi l_q}$, where
$\frac{8R}{\pi l_q}$ is the normalized phase space area taken up by disk $q$
if it is not shadowed, and $\hat{\eta}_q$ is a pruning factor;
$0 \leq \hat{\eta}_q \leq 1$. We thus write
\begin{equation}
\hat{Z}(k)=
1-\sum_{q} \hat{\eta}_q\frac{8R}{\pi l_q}e^{-ikT_q}  \ \ .
 \label{eqn:Zhat2}
\end{equation}

The direction $(1,0)$ is always free if  $2R<1$. This means that every
disk \( ( {\tiny \begin{array}{c} n \\ 1 \end{array} } )  \)
is accessable and takes up a
normalized phase space area
\begin{equation}
a_{q=
{\tiny  \left( \begin{array}{c} n \\ 1 \end{array} \right)} }(0)
\approx \frac{2}{\pi} (2R+1/2R-2)\frac{2}{n(n-1)(n-2)}
\label{eqn:aqtail}
\end{equation}

Let's now consider the zeta function $Z(k)$.
First we will neglect the factors with $m>0$ in \EqRef{Zcl}.
Given a periodic orbit $\overline{s_1 s_2 \ldots s_n}$ we
make the following assumtion ({\em assumtion B}):
\begin{enumerate}
\item The accessable values for $s_i$ does not depend on $s_j$, $j \neq i$.
\item The stability eigenvalues are given by
      $\Lambda_{\overline{s_1 \ldots s_n}}=
\prod_{i}\Lambda_{\overline{s_i}}$
\item The periods are given by $T_{\overline{s_1 \ldots s_n}}=
\sum_i T_{\overline{s_i}}$
\end{enumerate}
This assumption is not at all true, although it is realistic if all
$T_{\overline{s}}$'s are small.
We introduce it as a working assumption
to shed some light on the nature of the BER approximation.

Condition 1 means that we can make a curvature expansion \cite{AAC}.
Conditions 2
and 3 means that all curvatures disappear and what remains is the fundamental
part \begin{equation}
Z(k)=1-\sum_{s}
\frac{\eta_{\overline{s}}}{\Lambda_{\overline{s}}}
e^{-ikT_{\overline{s}}} \ \ .
 \label{eqn:Zfund}
 \end{equation}
 $\eta_{\overline{s}}$ is a
pruning
factor which is $=1$ if cycle $\overline{s}$ exist, and zero otherwise.
Since we have assumed that $T_{\overline{s=(q,g)}}$ is a
function of $q$ only
($\equiv T_q$) we
sum over $g \in C_{4,v}$ keeping
\( q=( {\tiny \begin{array}{c} n_x \\ n_y \end{array} } )  \)
fixed and get
\begin{equation}
Z(k)=1-\sum_{q\in \omega }
\eta_q\frac{R}{2l_q}{\left[(1+\sqrt{2})(1+cos\phi_q)+sin\phi_q \right] }
e^{-ikT_q} \ \ .
 \label{eqn:Zfundexp}
\end{equation}
The angle $\phi_q$ is the angle between the horisontal axis
and the direction vector $(n_x,n_y)$.
$\eta_q$ takes on discrete values and is $=1$ if all
$g's$ are realised. In fig. 2 we
compare the prefactors (in arbitrary units) of eqs. \EqRef{Zhat2} and
\EqRef{Zfundexp} provided that $\eta_q=\hat{\eta}_q$.
The similarity with eq. \EqRef{Zfund} is obvious and we may deduce that
Assumptions A and B are similar. The reason for the agreement is that
the available phase space region in eq. \EqRef{Zhat}
is extrapolated from the local stabilities in eq. \EqRef{Zfundexp}.
This should work well as long as there is no shadowing (=obstructing)
disks in between.


We will now see that
eq. \EqRef{Zfundexp} fails completely
when pruning becomes essential, that is when
$\hat{\eta}_q \ll 1$. This is e.\ g.\ the case far out
in the free directions.
Consider the free direction $(1,0)$. We see from \EqRef{aqtail} that
\( a_{q=( {\tiny \begin{array}{c} n \\ 1  \end{array} })}(0)  \) decays as
$\sim 1/n^3$.
There is only one period-one cycle with
\( q=( {\tiny \begin{array}{c} n \\ 1 \end{array} })  \),
and it is the one
having $g=\sigma_{-}$ and
the corresponding prefactors $1/\Lambda_{\overline{(q,\sigma_{-})}}$
decays as
$1/n^2$!

The reason for the discrepancy is the fake assumption B.
An apparent example of
violation of B1 is the following. The cycle $\overline{s}$ where $s=(q,g=E)$
is pruned but the same symbol $s$ may very well appear in longer cycles.

As we have already said, there is no well
defined fundamental part of the expanded zeta function.
Indeed, by our
discussion we are led to suggesting that the BER approximation provides
a natural
generalization of the fundamental part, with the extra advantage of
preserving  unitarity: $k_0=0$ is by construction a zero of the unweighted
zeta function $\hat{Z}(k)$.

We have seen that the problem of establishing a fundamental part of the
zeta function using periodic orbits is
connected with the marginal orbits
which we have pruned by excluding the marginal phase
and the ackumulation
of hyperbolic orbits towards it.
The BER approximation
tells us that there may be simpler means of exploring the available
phase space for a certain symbol than letting periodic orbits explore it.

\subsection{Other representations, other weigths in the
BER approximation}

It is straightforward to study other than the symmetric representation
$A_1$ in the BER approximation.
In this approximation we only consider one single disk bounce.
Since we have decided that the trajectory goes out into the first
octant it suffices to determine what octant it came from in order to
determine the group element $g_{s}$
 associated with
this particular bounce.
That is, we divide the phase space region $\Omega_q$ into eight
regions $\Omega_s$ where $s=(q,g)$ and $g \in C_{4,v}$.
Some of these may of course
be empty.

The group character associated with the whole interval (one bounce +
the trajectory to the next disk) is then (cf, \EqRef{symmfac})
\begin{equation}
\chi_r(g_s)=\chi_r(g) \chi_r(\sigma_{-})^{n_x+n_y}  \ \ .
\end{equation}

The zeta function may then
be written
\begin{equation}
\hat{Z}_{r}(k)
\approx 1-\sum_{s=(q,g)}\chi_r(g_s) a_{s}(0) e^{-ikT_s} \ \ ,
 \label{eqn:Zhatcl}
\end{equation}
where in $a_{s}(0)$ is defined by  \EqRef{aqdef} but the integral extends
only over
 $\Omega_s$.

It is also straightforward to use other weigths $w$
in the BER approximation for
the Sinai billiard. All weights we are going to study in this paper belong to
the family $w=\Abs{\Lambda(\phi,\alpha)}^{\tau}$.
The corresponding family of zeta functions then reads
\begin{equation}
\hat{Z}_{\tau,r}(k)
\approx 1-\sum_{s} \chi_r(g_s)a_{s}(\tau) e^{-ikT_s} \ \ ,
 \label{eqn:Zhattau}
\end{equation}
where
\begin{equation}
a_s(\tau)=\frac{2}{\pi}\int_{\Omega_s}
\Abs{\Lambda(\phi,\alpha)}^{\tau}d\phi
\;  d(sin\; \alpha)
\label{eqn:asdef}
\end{equation}
\begin{equation}
\Lambda(\phi,\alpha)\approx \frac{2l_s}{Rcos \; \alpha}  \ \ .
\nonumber
\end{equation}
As $\Lambda(\phi,\alpha)$ is not constant over $\Omega_s$,
c.f. eq. \EqRef{hatZw}, we have averaged it over $\Omega_s$.

\section{Computations and results}

\begin{figure}
\epsffile{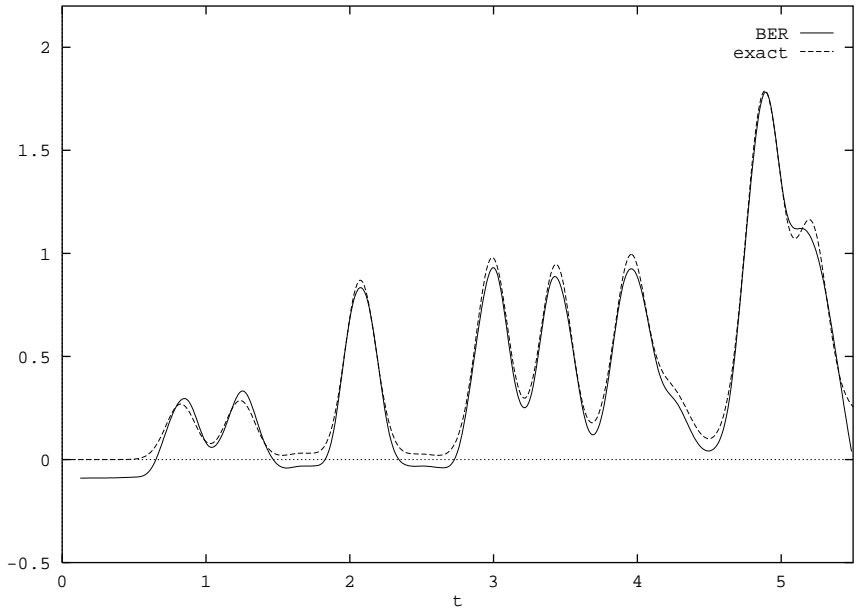}
\caption{The trace of the unweighted evolution operator.
The parameters are $R=0.1$, $r=A_1$, $\sigma=0.1$}
\end{figure}

We now turn to explicit computations. The aim is to check the validity
of the BER approximation as outlined in section 3.5, meaning that we restrict
ourselves to the family of weights $w=|\Lambda|^\tau $.

The strategy is the following.
First, we compute the trace in the BER approximation,
\begin{equation}
tr \hat{\cal L}_{\tau,r,\sigma}^t = \frac{1}{2\pi i}
\int_{-\infty}^{\infty} e^{ikt}
\frac{\hat{Z}_{\tau,r}'(k)}{\hat{Z}_{\tau,r}(k)}
e^{-k^2\sigma^2/2} dk   \ \ .
\label{eqn:traceBER}
\end{equation}

Then, in order to check this result, we
compute the 'exact' trace, by means of the trace formula
\begin{equation}
tr {\cal L}_{\tau,r,\sigma}^t  =
\sum_p T_p \sum_{n=1}^{\infty} \chi_r(g_p)^n |\Lambda_p|^{n\tau}
\frac{e^{(t-nT_p)^2/2\sigma^2}}{\Abs{(1-\Lambda_p^n)(1-1/\Lambda_p^n)}}  \ \ ,
\label{eqn:exact}
\end{equation}
To this
end we need the periodic orbits.
We use gaussian smearing in order to avoid the delta peaks.

The full trace of the full system (fig 1b) is obtained as a sum of traces
for all irreducible representations of $G=C_{4,v}$. As we will  not treat
two dimensional representations we will not perform the sum.

We mentioned before that orbits running along symmetry lines
has to be given a special treatment.
There are two such cycles in the Sinai billiard. We treat them in the
following way. One has symmetry $\sigma_{-}$. We simply exclude from
the trace formula \EqRef{exact} if
$\chi_r(\sigma_{-})=-1$.
The other one, having symmetry $\sigma_{/}$ is treated analogously.
This procedure
is not exact, c.\ f.\ \cite{L,CEsym}, but very accurate.

The computational task lies in determining the quantities
$a_s(\tau )$ as defined by eq. \EqRef{asdef}.

We do this in the simplest thinkable way by binning the
phase space corresponding to a disk $q$
and deleting the phase space shadowed by nearer disks.
More exactly, we bin the $\alpha$ variable, keeping track of the
limits in the $\phi$ direction.

The integrals will for obvious reasons diverge if $\tau>2$.

Of course, we only calculate  a finite number of $a_s$. Then
there are infinite tails in all free directions.
Let us for a minute restrict ourselves to the case $r=A_1$
and the (1,0) free direction.
Far out in the free directions only a small fraction
of disk \( q=( {\tiny \begin{array}{c} n \\ 1 \end{array} })  \)
is visible. Then $l_{s=(q,g)}$ is essentially independent of
$g$. If we sum over $g$ keeping $q$ fixt
\begin{equation}
\sum_{g} a_{s=(
{\tiny \left( \begin{array}{c} n \\ 1 \end{array} \right) }
 ,g) } (\tau)
\approx \frac{2}{\pi}(\frac{2n}{R})^{\tau}
\frac{(2/R-4)^{1-\tau/2}(1-2R)}{2-\tau}
\frac{1}{n^{1+\tau/2}(n-1)^{1-\tau/2}(n-2)^{1-\tau/2}}
\label{eqn:ovalong}
\end{equation}
We will not bore the reader with details about the other $r$'s and
free directions.
These infinite tails will make the zeta functions diverge in the upper
half $k$-plane. It is thus crucial that we find an analytical continuation
of the zeta function.

We also have to choose the lengths $l_s$ associated with
$\Omega_s$. We simply choose it to be the period of the cycle $\overline{s}$
regardless if it is pruned or not.
We choose the energy such that the lengths $l_s$ and the times $T_s$
coincides.

We will also calculate all periodic orbits up to some maximal period.
We thus generate symbol codes according to the scheme outlined
in the previous section. The orbits are then found by an extremum principle
according to ref. \cite{SS}. Last it is checked if each cycle is allowed
or pruned.

\subsection{The unweighted zeta function, $\tau=0$.}

We begin by considering the summation of the tails.
Consider disk
\( q=( {\tiny \begin{array}{c} n \\ 1 \end{array} })  \)
in the free direction $(1,0)$.
Only a small fraction of this disk is visible for large $n$. Then
$\Omega_{s=(q,g)}$ is nonempty only for $g=E$ and $g=\sigma_{-}$ and all
points lie on about the same distance $l_s=l_q$. Restricting ourselves to
$r=A_1$ we have to consider the following series
\begin{equation}
\sum_{n=N+1}^{\infty}\frac{2}{n(n-1)(n-2)}e^{-ikT_{q}}  \ \ .
\label{eqn:Zhattail}
\end{equation}
%
If we as an approximation put $l_{q=(n,1)} \approx n$ the
series reduces to a Fourier series in $z=exp(-ik)$.
This is easily summed since
\begin{equation}
\sum_{n=3}^{\infty}\frac{z^n}{n(n-1)(n-2)}=
-\frac{1}{2}(1-z)^2\log (1-z)+\frac{3}{4}(1-z)^2-(1-z)+\frac{1}{4} \ \ .
\end{equation}
The zeta function will thus have branch cuts along
$Re(k)=2\pi N$ and $Im(k)>0$.

There are similar contributions in all free direction.
The free direction $(x,y)$ will
induce branch cuts along $Re(k)=2\pi N/\sqrt{x^2+y^2}$ and $Im(k)>0$.

Close to the origin $k=0$ the zeta function behaves as
$Z(k)=c_1 (ik)+d_2 (ik)^2\log (ik) + c_2 (ik)^2 \ldots$,
$(r=A_1)$. Consequently the trace will have the asymptotic behaviour
\begin{equation}
tr \hat{\cal L}_{\tau=0,r=A_1,\sigma}^t \sim 1-C/t
\label{eqn:asymp}
\end{equation}
where $C$ is a positive constant, it may be related to the expectation
value of $p(\Delta)$ \cite{PDreson}.
The convergence of the trace formula towards unity is
usually reffered to as a periodic orbit sum rule \cite{Ozo}.

When considering other representations, $r$, some tail series will have
alternating signs. This will shift the branch cuts in the
real direction. These zeta functions will not have a root at the origin and
the traces goe asymptotically to zero.

We now turn to  the numerical computation of the trace.
When we calculate the Fourier transform we simply make the
integration along the the real axis, except at the origin which
we sneak just below.

\begin{figure}[p]
\epsffile{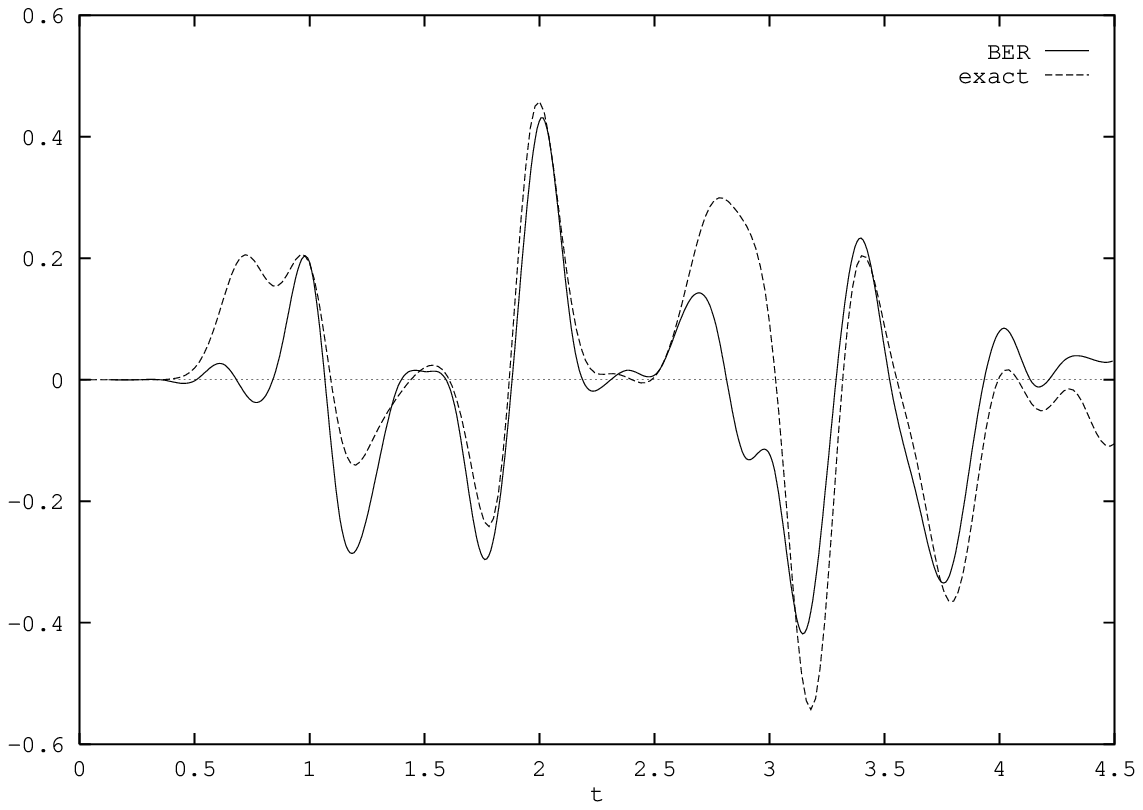}
{a) $R=0.2$, $r=B_2$, $\sigma=0.1$}
\end{figure}
\begin{figure}[p]
\epsffile{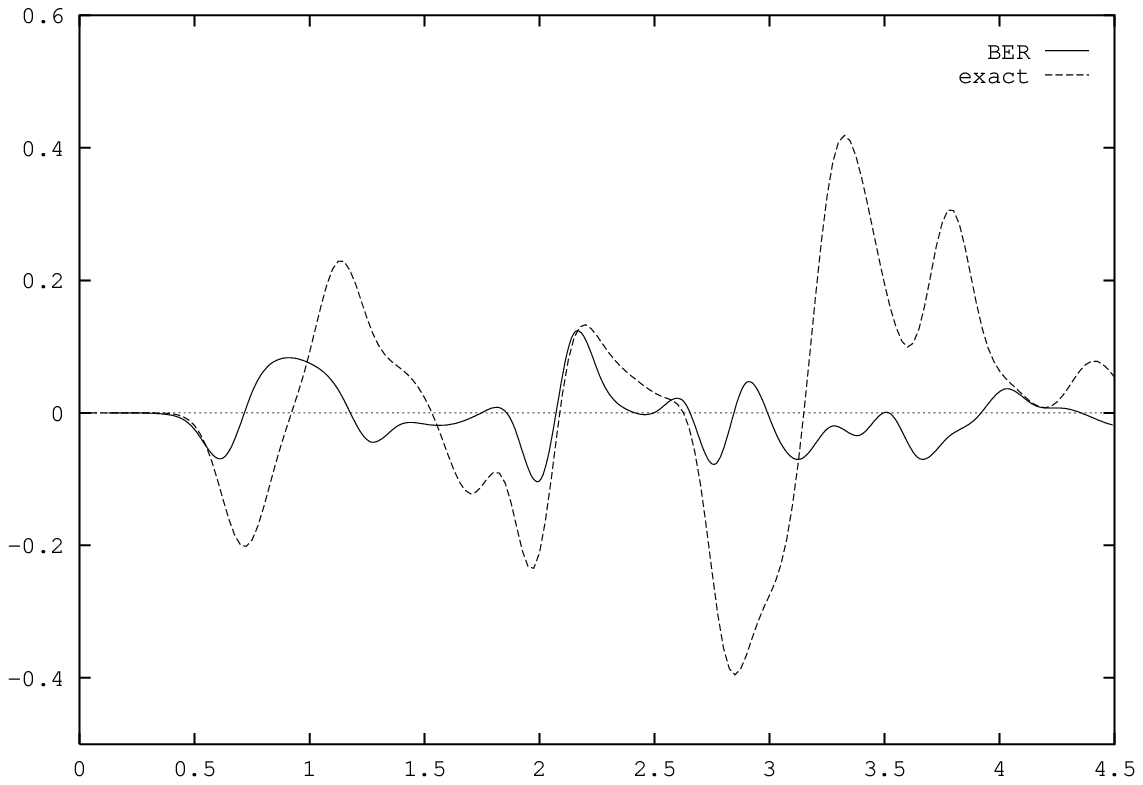}
{b) $R=0.2$, $r=A_2$, $\sigma=0.1$}
\end{figure}
\begin{figure}[p]
\epsffile{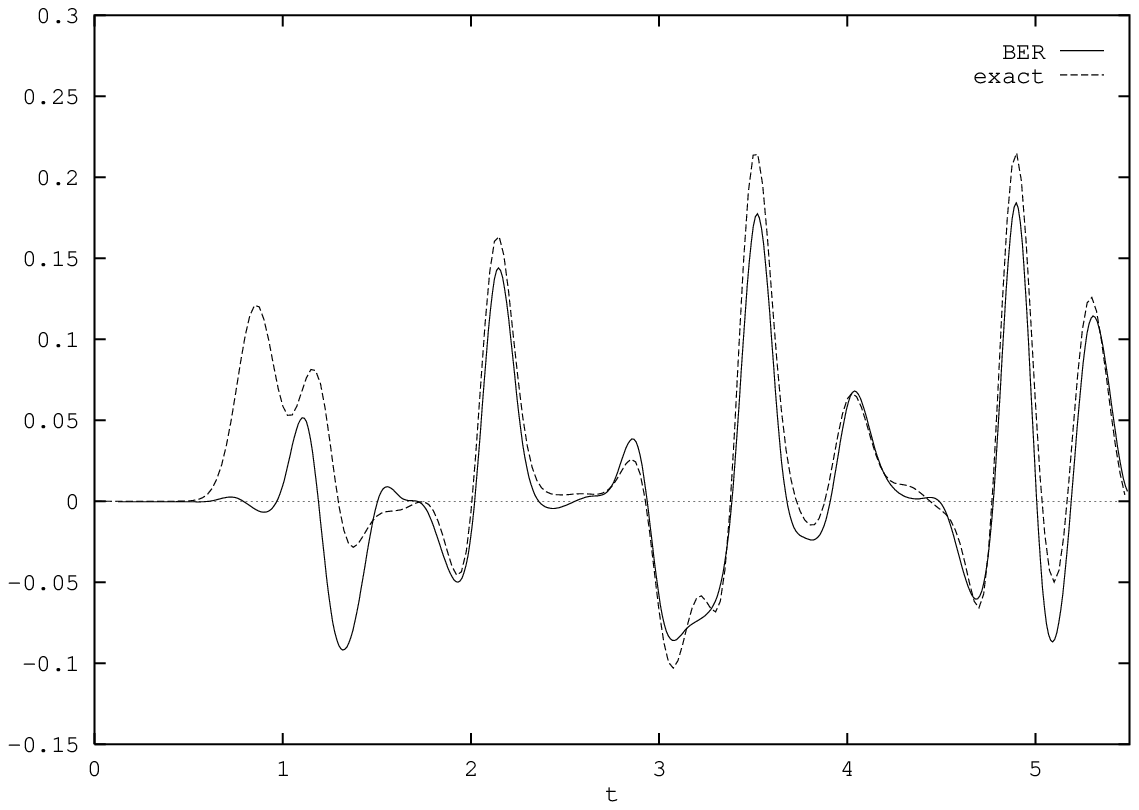}
{c) $R=0.1$, $r=B_2$, $\sigma=0.1$}
\end{figure}
\begin{figure}[p]
\epsffile{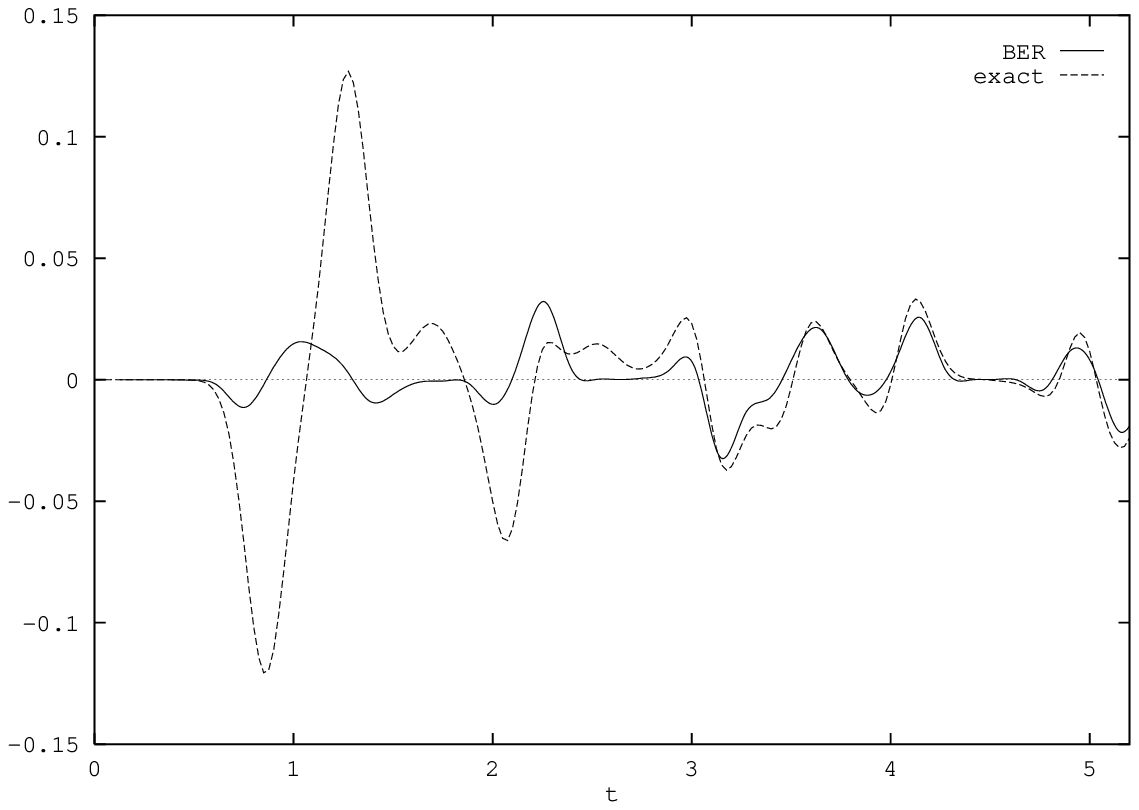}
{d) $R=0.1$, $r=A_2$, $\sigma=0.1$}
\caption{Trace of the unweighted evolution operator obtained from the
BER approximation and directly from periodic orbits}
\end{figure}

With the number of periodic orbits calculated (between 5000 and 8000
depending on $R$) we cannot check the asymptotic limit discussed above.
To do the numerical check we restrict ourselves to much shorter times $t$.
The result is seen in figs 3 and 4. The agreement is in most cases
very good. We note
\begin{itemize}
\item For small $t$ the agreement is less good or even lousy.
The exact result for small $t$ depends on a small number of cycles and
the BER theory requires an average effect from many cycles.
\item For fixed representation
$r$, the agreement is better for small radii $R$. This is
exactly what we expected.
\item For fixed radius $R$ the agreement is best for the symmetric
representation $A_1$ and worst for the anti symmetric $A_2$.
In the former case all $\Omega_s$ provides the same sign but otherwise
different phase space regions come with different signs
(provided by the group characters $\chi$).
It takes more periodic orbits in order to provide averages for this
finer partition.
\end{itemize}

We have seen that the trace is almost entirely governed by the branch cuts.
(There are a few isolated zeros but they are extremely sparse and have
no practical importance).
We are thus faced with an entirely different situation than for Axiom-A
systems for which the zeta function is entire and the zeros gives the
discrete spectrum of the operator.

We have not discussed in what sense the cuts may be interpreted
as a continous part of the spectra. If the trace, by means of the sum over
eigenvalues, is not mathematically well defined, we let it be
defined by means
of the corresponding periodic orbit sum.

\subsection{Topological entropy, $\tau=1$}

We now have to find the analytic continuation for sums of the form,
cf. eq. \EqRef{ovalong},
\begin{equation}
\sum_{n=N+1}^{\infty} \frac{e^{-ikcn}}{n^{3/2}}  \ \ .
\label{eqn:sven}
\end{equation}
This is just a power series in $exp(-ikc)$. We observe that
\begin{equation}
\sqrt{1-z}=1-\sum_{n=1}^{\infty}\frac{(2n-3)!!}{(2n)!!}z^n \ \ .
\end{equation}
Using Stirlings formula we find the following asymptotic expression
for the coefficient
\begin{equation}
\frac{(2n-3)!!}{(2n)!!} \sim \frac{1}{\sqrt{4\pi}}\frac{1}{n^{3/2}}
\end{equation}
enabling an summation of the \EqRef{sven}, provided of course that
$N$ is sufficiently big.
The zeta function still have cuts in the upper half plane.
However, there are isolated zeros far down in the lower half plane.
The determination of these does not depend very much on the summation
of the tails. Therefore we could take only the leading order of the tails
series into account above.

\begin{figure}
\epsffile{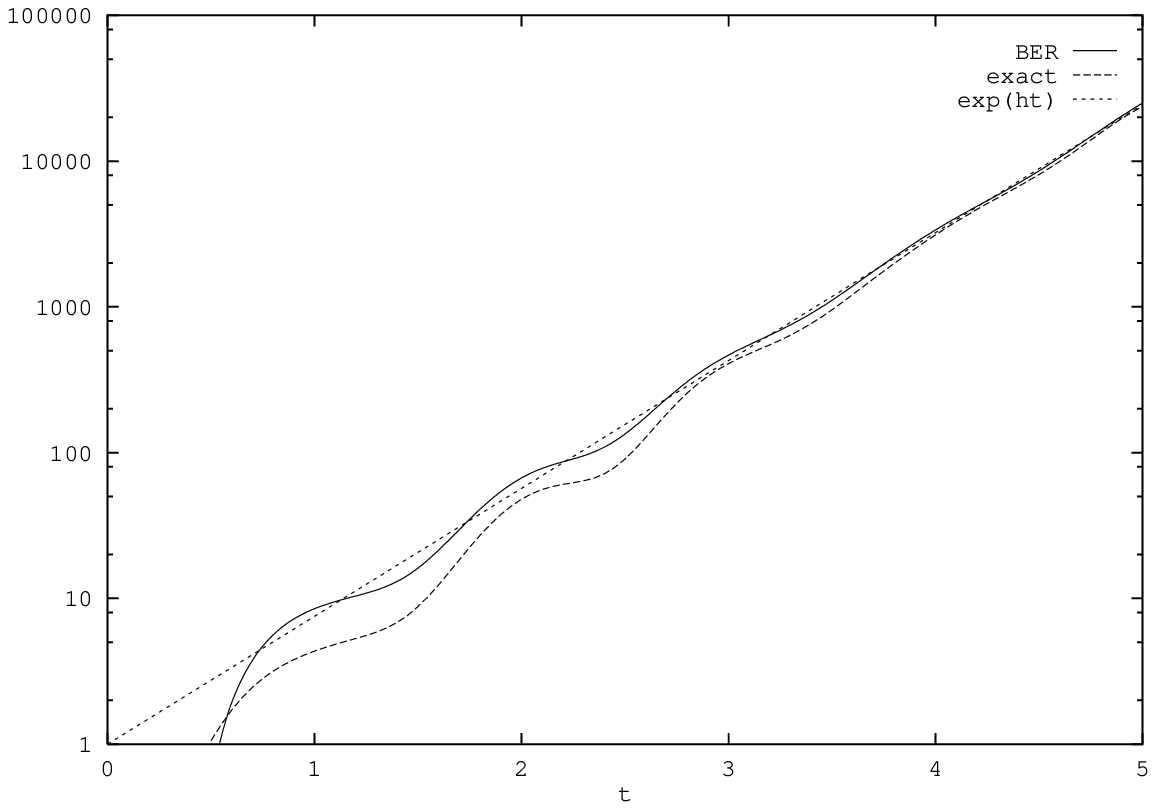}
{$R=0.1$, $r=A_1$, $\sigma=0.25$}
\caption{The trace of the topological evolution operator obtained from the
BER approximation and directly from periodic orbits}
\end{figure}

When preforming the integral \EqRef{traceBER} we let the contour go
below all singularities of the integrand, that is below all discrete
zeros $k_{\alpha}$ of $\hat{Z}$.
The sum over residues gives
\begin{equation}
tr \hat{\cal L}_{\tau=1,r=A_1,\sigma}^t \sim
\sum_{\alpha} e^{ik_\alpha t}
e^{-k_\alpha ^2\sigma^2/2} \ \ .
\label{eqn:toptrace}
\end{equation}
The result is seen in fig. 5. This result is again good for large $t$.

Of course, in the contour integration we have neglected the contribution
coming from the integration around the branch cuts cf ref. \cite{PDreson}.
This
should give an extra contribution $\sim -1/\sqrt{t}$ which is of course
overwhelmed by the exponential increase, but could explain the discrepancy for
small $t$ in fig. 5.

We have determined the topological entropy $h(R)$
for two different radii $R$. We find $h(0.2)=2.1346$ and
$h(0.1)=2.02024$. The entropy $h(R)$ tends to a finite limit
when $R \rightarrow 0$. It limiting value is obtained from the equation
\begin{equation}
1-\sum_{i,j \in Z, \; gcd(i,j)=1}e^{-h(0)\sqrt{i^2+j^2}}=0  \ \ .
\end{equation}
This is derived directly from eqs. \EqRef{Zhattau} and \EqRef{asdef} but is
indeed exact, thus demonstrating consistency of our approach. It allows
determination of $h(0)$ within a fraction of a second of computer time.
We find $h(0)=1.9133307629\ldots$. This value differs
slightly from Berry's estimate \cite{BerSin}
$h(0) \approx \sqrt{12/\pi }=1.95441$.

Considering the one parameter family of weights $|\Lambda |^\tau$, we see that
if $\tau >0$ the leading zeros lie below the branch cut and the trace
exhibit exponential behaviour. When $\tau =0$ the leading zeros coincides
with a branch point. When $\tau \leq 0$ the behaviour is entirely given by
the branch cuts, i.e. it is goverened by the tails in the free directions.
This is usually referred to as a phase transition \cite{phase}.
We have thus been able to give a nice analytical description of such a phase
transition
in terms of the zeta function; as the collision of the leading zero with a
singularity.

\section{Discussion}

The aim has been to show that the BER approximation works, and works well for
quite general weigths $w$. We have not aimed at calculating more physically
interesting properties, such as decay of correlations, fractal dimensions etc.
This is of course
a natural next step.

Regarding correlation functions one may, loosely, say that they behave
very much like the the trace of the evolution operator with unit weight.
This is easily seen if the eigenvalues $\exp (ik_nt)$ are isolated. Then
the correlation function $C(t)$ may (formally) be expanded as $C(t)=\sum_n c_n
\exp(ik_nt)$. The coefficients $c_n$ depend on the observables.
The trace $tr{\cal L}^t = \sum_n \exp(ik_nt)$ may therefore be thought of
as a archetype correlation function.
The velocity autocorrelation function for the Sinai billard (with
continous time) have been reported to decay as $1/t$ \cite{Frid}, which
is the same decay as we found for the trace.

The BER approximation is in principle applicable to a wide range of bound
ergodic Hamiltonian systems. The details have to be worked out for each
separate case. For the Sinai billiard it turned out to be particularly simple.
The reason for this is that we could make the approximation that the lengths
$l_s$ are constant over the phase space regions $\Omega_s$. This was because
the flow is suspended in the sense that liminar segments of arbitrary
lengths does not exist. For instance, there is no such length shorter than
$1-2R$. This has to do with the presence of marginally stable orbits.
A consequence of this is the periodicity of the cuts.

The situation is very different for e.g.\ the hyperbola billiard. In ref.
\cite{PDreson} we found evidence that the zeta function ($\tau =0$) for this
system has a discrete spectrum of zeros and a single cut along the positive
imaginary axis. We used cycle expansions to obtain this result, but since the
zeta function is divergent we had to do some ad hoc manipulations.
The expansion used exhibited some obvious similarities with the BER
approximation worked out in this paper. The expansion consisted almost
exclusively of cycles with only one laminar segment. To avoid
(the most serious)
divergence we selected out an infinite subsequence
ackumulating towards a pruned orbit and
 responsible for the
branch cut. This sequence has as a counterpart the tails in the free
directions of the Sinai billiard.

Perhaps the most powerful result in this article is the possibility of
making asymptotic statements, cf eq \EqRef{asymp},
about periodic orbits.
By varying the parameter $\tau$ one can obtain a wealth of such {\em periodic
orbit sum rules}. This asymptotics is essential for doing semiclassical
calculation on spectral fluctuations based on
Gutzwiller's trace formula \cite{Berry,PDreson}.
We have therfore outlined how to do periodic orbit theory without a single
periodic orbit.

\section*{Acknowledgements}

I would like to thank Cecilia Kozma for a useful hint.
This work was supported by the Swedish Natural Science
Research Council (NFR) under contract no. F-FU 06420-303.


\newcommand{\PR}[1]{{Phys.\ Rep.}\/ {\bf #1}}
\newcommand{\PRL}[1]{{Phys.\ Rev.\ Lett.}\/ {\bf #1}}
\newcommand{\PRA}[1]{{Phys.\ Rev.\ A}\/ {\bf #1}}
\newcommand{\PRD}[1]{{Phys.\ Rev.\ D}\/ {\bf #1}}
\newcommand{\PRE}[1]{{Phys.\ Rev.\ E}\/ {\bf #1}}
\newcommand{\JPA}[1]{{J.\ Phys.\ A}\/ {\bf #1}}
\newcommand{\JPB}[1]{{J.\ Phys.\ B}\/ {\bf #1}}
\newcommand{\JCP}[1]{{J.\ Chem.\ Phys.}\/ {\bf #1}}
\newcommand{\JPC}[1]{{J.\ Phys.\ Chem.}\/ {\bf #1}}
\newcommand{\JMP}[1]{{J.\ Math.\ Phys.}\/ {\bf #1}}
\newcommand{\JSP}[1]{{J.\ Stat..\ Phys.}\/ {\bf #1}}
\newcommand{\AP}[1]{{Ann.\ Phys.}\/ {\bf #1}}
\newcommand{\PLB}[1]{{Phys.\ Lett.\ B}\/ {\bf #1}}
\newcommand{\PLA}[1]{{Phys.\ Lett.\ A}\/ {\bf #1}}
\newcommand{\PD}[1]{{Physica D}\/ {\bf #1}}
\newcommand{\NPB}[1]{{Nucl.\ Phys.\ B}\/ {\bf #1}}
\newcommand{\INCB}[1]{{Il Nuov.\ Cim.\ B}\/ {\bf #1}}
\newcommand{\JETP}[1]{{Sov.\ Phys.\ JETP}\/ {\bf #1}}
\newcommand{\JETPL}[1]{{JETP Lett.\ }\/ {\bf #1}}
\newcommand{\RMS}[1]{{Russ.\ Math.\ Surv.}\/ {\bf #1}}
\newcommand{\USSR}[1]{{Math.\ USSR.\ Sb.}\/ {\bf #1}}
\newcommand{\PST}[1]{{Phys.\ Scripta T}\/ {\bf #1}}
\newcommand{\CM}[1]{{Cont.\ Math.}\/ {\bf #1}}
\newcommand{\JMPA}[1]{{J.\ Math.\ Pure Appl.}\/ {\bf #1}}
\newcommand{\CMP}[1]{{Comm.\ Math.\ Phys.}\/ {\bf #1}}
\newcommand{\PRS}[1]{{Proc.\ R.\ Soc. Lond.\ A}\/ {\bf #1}}

\end{document}